\documentclass[aps,12pt,superscriptaddress,amsfonts,amssymb,amsmath]{revtex4}
\pdfoutput=1

\usepackage{graphicx}
\usepackage{epsfig}
\usepackage{epstopdf}
\usepackage{natbib}
\usepackage{xcolor}

\usepackage{verbatim}
\usepackage{empheq}
\usepackage{dsfont}
\usepackage{caption}
\usepackage{subcaption}
\usepackage{tabulary}
\usepackage{nccmath}
\usepackage{flafter}
\usepackage{array}

\begin{document}

\title{Diffusion on assortative networks: from mean-field to agent-based, via Newman rewiring}

\author{L.\ Di Lucchio \footnote{Email address: Laura.DiLucchio@unibz.it}}
\affiliation{Free University of Bozen-Bolzano \\ Faculty of Engineering \\ I-39100 Bolzano, Italy}

\author{G.\ Modanese \footnote{Email address: Giovanni.Modanese@unibz.it}}
\affiliation{Free University of Bozen-Bolzano \\ Faculty of Engineering \\ I-39100 Bolzano, Italy}
\date{\today}

\linespread{0.9}

\begin{abstract}
In mathematical models of epidemic diffusion on networks based upon systems of differential equations, it is convenient to use the Heterogeneous Mean Field approximation (HMF) because it allows to write one single equation for all nodes of a certain degree $k$, each one virtually present with a probability given by the degree distribution $P(k)$. The two-point correlations between nodes are defined by the matrix $P(h|k)$, which can typically be uncorrelated, assortative or disassortative. After a brief review of this approach and of the results obtained within this approximation for the Bass diffusion model, in this work we look at the transition from the HMF approximation to the description of diffusion through the dynamics of single nodes, first still with differential equations, and then with agent-based models. For this purpose, one needs a method for the explicit construction of ensembles of random networks or scale-free networks having a pre-defined degree distribution (Configuration Model) and a method for rewiring these networks towards some desired or ``target'' degree correlations (Newman Rewiring). We describe Python-NetworkX codes implemented for the two methods in our recent work and compare some of the results obtained in the HMF approximation with the new results obtained with statistical ensembles of real networks, including the case of signed networks.

\ 


\end{abstract}

\maketitle

\section{Introduction}

Mathematical models of diffusion processes on networks are an important tool especially for social and economic sciences, as they allow to analyse and predict the spreading of innovation, information, physical or financial contagion among individuals or companies.

In this work we will mainly consider as working example the Bass model for innovation diffusion, which is of interest for marketing analysis and also concerning diffusion of social practices \cite{bass1969new,norton1987diffusion,mahajan1990determination,van2007new,goldenberg2009role,rogers2010diffusion,jiang2012generalized,schilling2019strategic}. The Bass model includes as special case the widely used SI (Susceptible-Infected) epidemiological model, and has also been employed for modelling the spreading of information and rumors \cite{moreno2003disease,keeling2005networks,nekovee2007theory}. This last application has taken advantage in recent times from the large amount of data collected through social media, while the classical Bass model has been typically applied to ``real'' marketing data regarding single customers or firms and institutions.

Most diffusion models were introduced decades ago in elementary form without a network of relations, as describing an idealized homogeneous population with all-to-all potential contacts. They have been historically defined in terms of systems of a few coupled ordinary differential equations. This means typically 2 or 3 equations for basic models, and up to one dozen if one introduces more complex evolution conditions like different categories of adopters (``innovators, laggards'', etc.), or different product generations, or adoption rules more complex than simple contagion, based for instance on utility functionals and the so-called ``4P marketing mix'' Product-Price-Place-Promotion \cite{mccarthy1979basic,mesak1996incorporating,pinto2022marketing}.

With these relatively small equation systems it is possible to make some rigorous analysis also at the level of control and optimal control (\cite{oliinyk2018optimal} and refs.), in addition to look for exact solutions in special cases and to numerical (plus possibly stochastic) solutions in the other cases.

When we turn to consider systems of equations on networks, we are facing a few fundamental alternatives both at the level of the network itself and of the kind of equations. 

The network can be of the random Erd\"os-Renyi type (with exponential tail in the degree distribution) or scale-free (with a ``fat'' power tail, and thus with some very large hubs). 

The differential equations can be referred to the adoption or contagion of singles nodes, in which case there will be successive approximations of first, second etc.\ ``closure'' using the full network adjacency matrix \cite{newman2010networks,gleeson2011high,porter2016dynamical}; or they can be heterogeneous mean field (HMF) equations referred to the nodes' degrees, where the properties of the network are summarized through its statistical functions $P(k)$ (degree distribution) and $P(h|k)$ (degree correlations; see Sect.\ \ref{stat_descr}). In the first alternative, one will have a system of $\sim N$ equations, where $N$ is the number of nodes; in the second, $\sim n$ equations, where $n$ is the maximum degree. We recall that in a scale-free network with degree distribution $P(k) \sim k^{-\gamma}$, $N$ and $n$ are connected through the relation $n^{\gamma-1}\sim N$ (see Sect.\ \ref{stat_descr}); for example, for Barabasi-Albert networks we have $\gamma=3$ and $n\sim \sqrt{N}$.

For large scale-free networks the mean-field approximation is clearly convenient, also because the information about the network is in many cases available only in statistical, aggregated form. It is clear that since the HMF description regards as equivalent all nodes of the same degree, it will miss some details of the real diffusion processes. We shall mention this issue in several occasions in the following. Nevertheless, the HMF approximation allows to obtain important analytical results by making recourse to general techniques of statistical mechanics and theorems about linearized differential equation systems, especially concerning the epidemic threshold in scale-free networks \cite{pastor2001epidemic,pastor2002epidemic,boguna2002epidemic,boguna2003epidemic,boguna2003absence,boguna2004cut,barthelemy2004velocity,barthelemy2005dynamical,pastor2007evolution,pastor2015epidemic}.

In Sect.\ \ref{building} we shall describe the concrete realization of statistical ensembles of scale-free networks through techniques such as the Configuration Model and Newman rewiring, giving some new recipes implemented with Python-NetworkX codes. This kind of concrete network realizations lead to a better understanding of some conceptual properties like degree assortativity and its relation to the epidemic threshold (Sect.\ \ref{stat_descr}). Another important use of network realizations is for solving diffusion equations based on the single nodes, demonstrating deviations from the HMF approximation, and also for running agent-based simulations. In this way, we can make contact with another popular approach to diffusion studies, namely the technique of cellular automata (see \cite{kiesling2012agent} and refs.), which is much more flexible than differential equations because it allows to equip the agents with all sorts of complex behavior, but which has been traditionally studied only on simple regular lattices or random networks.

While in Sect.\ \ref{random-hubs} we only give a brief example of the results obtained with assortative rewiring, referring for more details to \cite{di2023generation}, in Sect.\ \ref{dis_rew_analysis} we describe some new results about disassortative rewiring based on target correlation matrices built with the Porto-Weber method. In this method, the correlation matrix $P(h|k)$ is re-constructed from an assigned decreasing $\bar{k}_{nn}(k)$ function ($\bar{k}_{nn}(k)$, the average nearest-neighbor degree function, is more frequently derived from a given $P(h|k)$ by summation of $hP(h|k)$ on $h$). In addition, we give examples of maximally disassortative networks built with a technique different from Newman rewiring, in which at any rewiring step one computes the variation $\Delta r$ of the assortativity coefficient through a ``local'' formula containing only the degrees of the four nodes involved in the rewiring. The rewiring step is then accepted or rejected with a criterion of positive/negative $\Delta r$, depending on whether a maximally assortative/disassortative network is desired. It is possible to define formally a temperature and thus assign a small return probability also to unfavourable changes, like in a familiar Metropolis-Monte Carlo algorithm. 

The maximally disassortative networks obtained with $k_{min}>1$ (in our example, $k_{min}=2$) are not fragmented like those with $k_{min}=1$ previously studied in \cite{moreno2003disease,bertotti2020network}. By analyzing their $\bar{k}_{nn}$ function, it is possible to gain useful information about the connections of their hubs. In Sect.\ \ref{relation} we recall a general relation between the variation $\Delta r$ in a degree-conserving rewiring and the variation $\Delta K$ in the same rewiring \cite{bertotti2020network}. $K$ is the network average of the degrees of the nearest neighbors of a node of any degree, and it is known to be related to the epidemic threshold (Sect.\ \ref{stat_descr}).

In Sect.\ \ref{maximally} we describe the main features of maximally disassortative networks obtained with a rewiring based on the criterion $\Delta r<0$. Sect.\ \ref{abm} summarizes our agent-based simulations performed with NetLogo, including the case of signed networks (model of Mueller-Ramkumar). Finally, Sect.\ \ref{conc} contains our conclusions, with a recapitulation of results about the peak diffusion times in dependence on the degree correlations.

\section{The statistical description of networks}
\label{stat_descr}

In this Section we recall some basic concepts of the statistical theory of networks, in order to place the following results in a more general context and to make them more easily accessible for the wide community of scientists who employ networks for purposes of modelling and data analysis.

\subsection{Degree distribution and correlations}

If we look at a network in abstract, or even as a practical data structure, we find that it is completely described by a list of the nodes and all their links, or equivalently by the adjacency matrix $A_{ij}$ \cite{newman2010networks,barabasi2016network}. By definition, $A_{ij}$ is equal to one if the nodes $i$ and $j$ are connected, and zero if they are not. However, for very large networks it is convenient to define some statistical descriptors, of which the most important are:

(1) The degree distribution $P(k)$, giving the probability that a node randomly chosen in the network has degree $k$. $k$ ranges in principle from 1 to a maximum degree $n$. The degree distribution is normalized to 1, see eq.\ \eqref{normalizations}.

(2) The two-point degree correlation function $P(h|k)$, giving the conditional probability that a randomly chosen node, if it has degree $k$, is connected to a node of degree $h$. This function is normalized to 1 in the first argument, see eq.\ \eqref{normalizations}.

The normalization relations are the following:

\begin{equation}
    \sum_{k=1}^n P(k)=1; \qquad \sum_{h=1}^n P(h|k)=1, \ \ \forall k=1,\ldots,n
\label{normalizations}
\end{equation}

Additionally, the following ``closure relation'' involving the degree distribution and the two-point degree correlation function must be satisfied, for consistency of the probabilistic description:

\begin{equation}
    hP(k|h)P(h)=kP(h|k)P(k), \ \ \forall h,k=1,\ldots,n
\label{closure}
\end{equation}

Many software packages for network analysis allow to compute these two quantities with simple commands starting from a ``network object'' $G$. For example, in NetworkX one can use the command \texttt{nx.degree\_histogram(G)} and then divide by the number of nodes, in order to obtain the degree distribution. The degree correlation matrix can be obtained with \texttt{nx.degree\_mixing\_matrix(G)}. A minor technical problem occurring with this correlation matrix of NetworkX is that if in a concrete network there are no nodes of degree $\hat{k}$, then the corresponding column $P(h|\hat{k})$ is absent. This can be avoided using the \texttt{mapping} command (see NetworkX documentation).

Note that the probabilities $P(k)$ and $P(h|k)$ do not contain information about other network metrics (path length, centrality, clustering...) and about network topology in general. Small changes in a network which are statistically irrelevant and do not substantially affect $P(k)$ and $P(h|k)$ can sometimes have a large impact on topology. Further note that higher-order correlations can be quite important in a network. Ref.\ \cite{bertotti2019configuration} describes an example of ``reconstruction'' via rewiring of a Barabasi-Albert-1 (BA1) network from its two-point correlations. The reconstructed network contains many disconnected triples absent in the true BA1 network; this happens because the rewiring procedure has no information about a three-point correlation like $P(1|2|1)$.

\begin{figure}[ht] 
\centering 
\includegraphics[width=0.6\columnwidth]{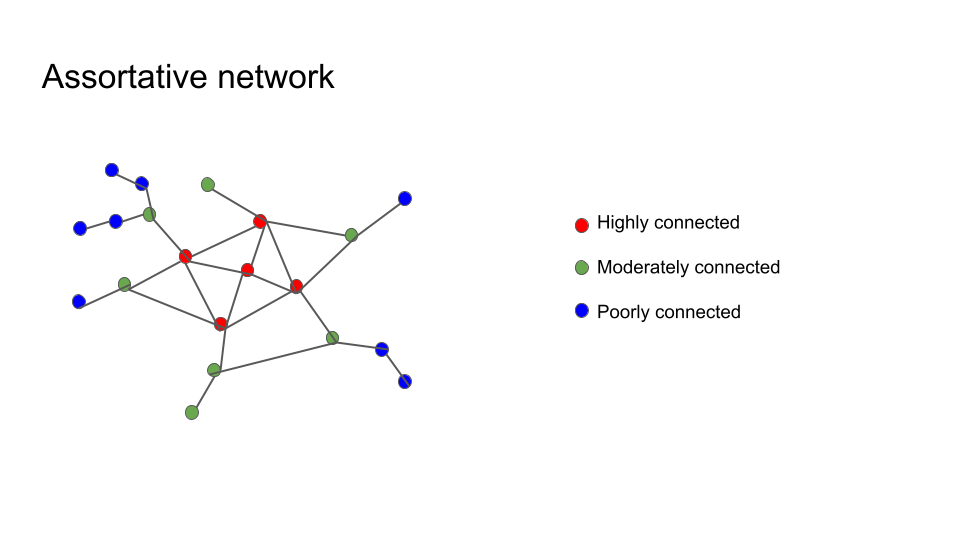}
\includegraphics[width=0.6\columnwidth]{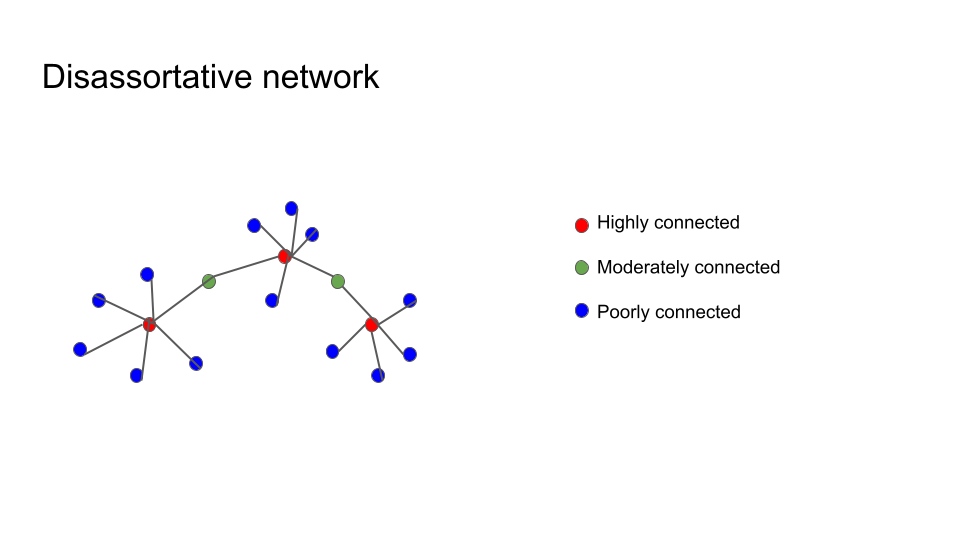}
\caption{
Concept of assortative and disassortative networks. In an assortative network, nodes tend to be connected to other nodes of similar size. Therefore, the hubs in the network tend to connect directly to other hubs and to connect to small nodes only through middle-size intermediate nodes. In a disassortative network, on the opposite, nodes of a certain size tend to be connected to nodes of a different size. This implies that the hubs tend to connect directly to small nodes, and to connect to other hubs only through small-size or possibly middle-size intermediate nodes.
}
\label{fig-ass-dis}
\end{figure}

Nevertheless, $P(k)$ and $P(h|k)$ are fundamental for the statistical description of networks. Their typical graphical representations are for $P(h)$ an histogram, for $P(h|k)$ a 2D plot. Contracted quantities computed with $P(k)$ are the momenta $\langle k \rangle=\sum_k kP(k)$, $\langle k^2 \rangle =\sum_k k^2P(k)$, etc. To contract $P(h|k)$ one can sum over $h$ and obtain a 1D plot of the so-called $\bar{k}_{nn}$ function, $\bar{k}_{nn}(k)=\sum_{k=1}^n hP(h|k)$
(average degree of the first neighbors of a node as a function of the degree of the node). When $\bar{k}_{nn}(k)$ is increasing, we normally have an assortative network, with $r>0$; when $\bar{k}_{nn}(k)$ is decreasing, a disassortative network, with $r<0$ (see Fig.\ \ref{fig-ass-dis}).

The definition of the Newman assortativity coefficient is the following:
\begin{equation}
    r = \frac{1}{\sigma_q^2}\sum_{j,k=0}^{n-1}jk(e_{jk}-q_{j}q_{k})
\end{equation}
Here $e_{jk}$ is the probability that a randomly chosen link in the network joins nodes with excess degrees $j$ and $k$, i.e., nodes which have $j$ and $k$ links in addition to the one we are considering. $q_j$ represents the normalized distribution of these excess degrees; it can be obtained as $q_j=\sum_{k=0}^{n-1} e_{jk}$ and is related to $P(k)$. Finally, $\sigma_q^2=\sum_{k=0}^{n-1}k^2q_k-(\sum_{k=0}^{n-1}kq_k)^2$.

A network is said to be uncorrelated, with regard to the degrees, if the correlation function $P(h|k)$ is independent from $k$ and takes the form $P(h|k)=hP(h)/\langle k \rangle$, corresponding to a constant $\bar{k}_{nn}$ function. An uncorrelated network has vanishing Newman coefficient $r$, but the converse in not generally true.
Indeed, more complex situations concerning degree correlations exist, e.g.\ in Barabasi-Albert networks. These networks have $r \simeq 0$, but using their $P(h|k)$ as recently given by Fotohui-Rabbat \cite{fotouhi2013degree} to compute their $\bar{k}_{nn}$ function, one sees they are not uncorrelated but disassortative at small degrees and assortative at large degrees \cite{bertotti2019bass}.
This means that, statistically, small nodes are likely to be attached to larger nodes, while hubs tend to connect quite directly to other hubs.

For mean-field modeling of diffusion on assortative networks it is necessary to introduce assortative correlation matrices. The simplest one is the correlation matrix by Vazquez-Weigt
\begin{equation}
    P(h|k)=(1-r)\frac{hP(h)}{\langle k \rangle} + r \delta_{hk}
\label{VW-matrix}
\end{equation}
whose $\bar{k}_{nn}$ function is the straight line
\begin{equation}
 (1-r)\frac{\langle k^2 \rangle}{\langle k \rangle} + r k   
\end{equation}
Other assortative matrices have been proposed in \cite{bertotti2016bass} and used for rewiring in \cite{di2023generation}. For modeling diffusion on disassortative networks, one can use correlation matrices built according to a recipe by Newman \cite{newman2003mixing}, involving the probability $e_{ij}$,  or a recipe by Porto and Weber \cite{weber2007generation}, see Sect.\ \ref{bass-model}.

\subsection{Diffusion dynamics in mean-field}

As mentioned, the knowledge of $P(h|k)$ is needed for writing the diffusion equations in HMF (Heterogeneous Mean Field) approximation.
This means having one equation for each degree, i.e., all nodes with the same degree are supposed to have the same behavior. For the Bass model the equations are 
\begin{equation}
    \frac{dG_i(t)}{dt} = [1-G_i(t)] \left[  p+iq \sum_{h=1}^n P(h|i)G_h(t)  \right], \qquad i=1,...,n
\end{equation}
where $G_i(t)$ is the fraction of potential adopters with $i$ links that at time $t$ have actually adopted.

The assumption that all nodes of the same degree have similar behavior is clearly false in general; consider e.g.\ a low degree node at the end of a chain, vs.\ some node with the same degree directly connected to a hub. In a real epidemic process, the first is infected later than the second. This will emerge from equations of single nodes. The HMF approximation essentially gives only averaged results.

In order to find a relation between maximum degree $n$ and network size $N$, we recall the Dorogotsev-Mendez statistical argument, valid for large scale-free networks. Assume there is probability 1 to have one node with degree $>n$. This defines $n$ as follows:
\begin{equation}
    \int_n^{+\infty} P(k)dk=\frac{1}{N}
\end{equation}
With $P(k)=c_{n,\gamma} k^{-\gamma}$ ($c_{n,\gamma}$ is a normalization constant of magnitude order 1, depending on $n$ and $\gamma$), we obtain $\frac{\gamma-1}{c_{n,\gamma}}n^{\gamma-1}=N$.
For example, for BA networks, which have $\gamma=3$, we obtain $n\simeq \sqrt{N}$, implying that with a system of e.g.\ 100 ordinary differential equations we can evaluate in HMF approximation the diffusion on a network with $N=10^4$ nodes. For computing diffusion on each node one should solve, in first closure approximation \cite{porter2016dynamical}, 10000 ordinary differential equations, which is barely possible with a normal machine. After that, a statistical analysis of results would be in any case necessary. A procedure of this kind has been executed, with small networks, in \cite{bertotti2019bass,bertotti2021comparison} and with larger networks in \cite{di2023generation}.

The HMF approximation has been much studied analytically for epidemic models of the SI type (Susceptible-Infected). 
The well-known results by Boguna, Vespignani et al.\ cited above concern mainly the (absence of) an epidemic threshold when the epidemics propagates on a scale-free network. Absence of a threshold means that no matter how small the contagion probability, the epidemics will eventually propagate to a finite part of the network. In the proof, the effect of correlations is only partially taken into account, i.e., with strong additional assumptions. With real assortative matrices one can see more complex behavior \cite{bertotti2021comparison}. For this reason, in certain cases it is also worth studying diffusion numerically through real implementations of networks with assortative correlation matrices (Sect.\ \ref{building}).

Is it possible, given a properly normalized $\bar{k}_{nn}$ function, to compute a correlation matrix which returns that $\bar{k}_{nn}$ upon contraction on $h$? Porto and Weber have proposed a method for this construction \cite{weber2007generation}, which has been used in \cite{weber2007generation} and \cite{silva2019spectral}. However, while the contraction of $P(h|k)$ gives a unique result, the correspondence $\bar{k}_{nn}(k) \to P(h|k)$ is not univocal. For ex., we have shown \cite{bertotti2021diagonal} that if one starts from the $\bar{k}_{nn}$ of the Vazquez-Weigt matrix, eq.\ (\ref{VW-matrix}), and then applies the Porto-Weber method to this function, the resulting correlation matrix is quite different from $P^{Vaz-Wei}$, and in particular the associated  connectivity matrix has a different eigenvalue spectrum. We recall that in general the connectivity matrix is defined as $C_{kh}=kP(h|k)$, and the largest eigenvalue $\Lambda^{max}$ of this matrix is equal to the reciprocal of the epidemic threshold $\lambda_c$ for the SI model in mean-field approximation. This means that if we write the mean-field equations as
\begin{equation}
    \frac{d\rho_k}{dt} = -\rho_k + (1-\rho_k) \lambda \sum_{h=1}^n k P(h|k) \rho_h, \qquad k=1, \ldots ,n
\end{equation}
where $\rho_k$ is the fraction of infected nodes with degree $k$, then $\lambda_c$ separates two different spreading scenarios: if $\lambda>\lambda_c$ the system reaches a final state in which a fraction of the population is infected; if $\lambda<\lambda_c$ the contagion dies out exponentially in time.

It can be shown \cite{bertotti2021diagonal} that for the Vazquez-Weigt connectivity matrix the largest eigenvalue is equal to $nr$. It follows, in particular, that when $r$ is small the epidemic threshold $\lambda_c$ is definitely greater than zero for small networks, but goes quickly to zero when $n \to \infty$. The convergence is much faster than for other correlation matrices, for which the largest eigenvalue typically grows as a root of $n$ or even as $\ln(n)$, when the scale-free exponent $\gamma$ is equal to 3. It was therefore concluded in \cite{bertotti2021diagonal} that adding even a small amount of diagonal degree correlations to an uncorrelated network leads, for large networks and in mean-field approximation, to a fast vanishing of the epidemic threshold.

\subsection{Bass model in HMF approximation}
\label{bass-model}

In the SI model, diffusion must be started by some seed adopters, and this involves arbitrary choices concerning the number of these adopters and their degrees.
In the Bass model, on the contrary, diffusion can start from a simple ``standard'' initial condition where the fraction of adopters of all degrees is zero. 
For the Bass model we can thus define and compute the time of the diffusion peak, or other diffusion-related times, in dependence on the kind of network and not on the initial conditions. In order to properly compare different networks, however, it is necessary to normalize the $q$-coefficient of the model to their average connectivity
\cite{bertotti2019evaluation}.

In this way it was found that diffusion is fastest on BA networks \cite{bertotti2019bass}. More exactly, for BA1 networks (with $k_{min}=1$) the peak time is about 88\% of the peak time for uncorrelated networks with the same scale-free exponent. For BA2 and BA3 networks the ratio is about 93\% and 96\%. For disassortative networks the peak time is slightly larger than for uncorrelated networks (about 104\% with correlation matrices from \cite{newman2003mixing}). For assortative networks the peak time is larger still, with correlations matrices from \cite{bertotti2016bass}.

We shall see that all these results depend strongly on the use of the mean-field approximation. The outcome is different if one builds real assortative scale-free networks with the configuration model plus Newman rewiring as explained in the following sections and implemented in our Python-NetworkX codes \cite{di2023generation}, and then solves the Bass equations for single nodes or performs agent-based simulations. In this case it turns out that the peak time is smaller for assortative than for uncorrelated networks, and that the difference is larger when the degree of the largest hubs present in the real network is larger, due to random fluctuations (while in the mean-field approximation all hubs are regarded as ``virtually present'', with a fixed small probability). 

The case of disassortative real networks has not been studied in detail yet, concerning the peak times. The reason is that the socio-economic networks for which it is interesting to simulate Bass diffusion are almost exclusively assortative. Nevertheless, it is possible to build real disassortative networks using the configuration model plus Newman rewiring with correlation matrices obtained through the Porto-Weber method. Some features of these networks are described below. Here we summarize for completeness the Porto-Weber recipe \cite{weber2007generation} that was implemented in our Python code already set up for the Newman rewiring. The conditional probability matrix $P(h\lvert k)$ is derived through a procedure named \texttt{Porto}, which reconstructs the expressions in \cite{weber2007generation}, namely:
\begin{equation}
    P(h\lvert k)=\frac{hP(h)}{\langle k \rangle}f(h,k)
\end{equation}
where $f(h,k)$ is the symmetric function
\begin{equation}
f(h,k)=1+\frac{[\bar{k}_{nn}(h)-k_{me}][\bar{k}_{nn}(k)-k_{me}]}{\langle k\bar{k}_{nn}\rangle_e-k_{me}^2}
\end{equation}
depending on the function $\bar{k}_{nn}$ and on two quantities defined as
\begin{equation}
k_{me}=\frac{\langle k^2 \rangle}{\langle k \rangle}   
\end{equation}
\begin{equation}
\langle k\bar{k}_{nn}\rangle_e=\sum_h \frac{hP(h)}{\langle k \rangle}h\bar{k}_{nn}(h)
\end{equation}
The conditional probability $P(h\lvert k)$ obtained in this way satisfies the normalization condition in $h$ and the network closure condition \eqref{closure}.

In the disassortative case, a suitable Ansatz for the $\bar{k}_{nn}$ function is $\bar{k}_{nn}\propto k^{-\beta}$, where $\beta$ is a positive number \cite{silva2019spectral}. In Sect.\ \ref{newman-rew} we describe the use of this Ansatz for the generation of ensembles of disassortative networks through Newman rewiring.

As already mentioned, the correlation matrix $P(h\lvert k)$ which yields the desired $\bar{k}_{nn}$ is not unique. In fact the definition of $\bar{k}_{nn}$ involves a summation and in principle any two correlation matrices $P^{(1)}$ and $P^{(2)}$ such that
\begin{equation}
\sum_h h [P^{(1)}(h\lvert k)- P^{(2)}(h\lvert k)]=0
\end{equation}
yield the same $\bar{k}_{nn}$. This was shown to be valid for any rearrangement of the $P(h\lvert k)$ in \cite{bertotti2021diagonal}. It is finally to be said that the Porto-Weber method can lead to negative elements for the values of the correlation matrix $P(h\lvert k)$ when $\bar{k}_{nn} \propto k^{\alpha}$, where $\alpha $ is a positive exponent such that $\alpha >0.4 $, and this was verified accurately with the Python code, e.g.\ in the case $\alpha=0.6$.

\section{Configuration model and assortative rewiring}

\subsection{Building ``real'' networks with (a) assigned degree distribution and (b) assigned correlations}
\label{building}

There are at least three reasons for which it is interesting to build statistical ensembles of networks having pre-defined degree distribution and correlations.

First, this is important for theoretical and foundational purposes, in order to understand whether networks of a certain kind really exist -- and not only their abstract degree distribution and correlation functions, satisfying the conditions of normalization and network closure, but otherwise quite arbitrary.

Second, for certain ``statistical experiments'' it is also crucial to build full ensembles of networks, which is just the outcome of the Newman rewiring procedure described below. For example, such ensembles were used for exploring the relation between the nearest-neighbor average degree $\bar{k}_{nn}(k)$ and the average clustering $\bar{C}(k)$ \cite{di2024bass}.

The third reason is that ``synthetic'' networks built through an algorithm are very convenient for testing models of dynamics and diffusion, or other network effects of interest for fundamental sciences and technology, both through the solution of differential equations and through simulations of agent-based models. 

The diffusion equations for the single nodes of a real network with adjacency matrix $A_{lj}$ are, in first closure \cite{porter2016dynamical}
\begin{equation}
    \frac{dX_l(t)}{dt}=\left( 1-X_l(t)\right) \left[ p+q\sum_{j=1}^N A_{lj} X_j(t) \right], \qquad l=1,\ldots,N
\end{equation}
where $X_l$ is the expectation value $\langle \xi_l \rangle$ of a random variable $\xi_l$ describing the non-adoption state ($\xi_l=0$) or the adoption state ($\xi_l=1$) of node $l$, taken over many stochastic evolutions of the system.

Condition (a) above (assigned degree distribution) can be realized exactly with the Configuration Model, which is essentially an extension of the method by Erd\"os-Renyi for the construction of uncorrelated random networks.
In the Configuration Model, given a degree sequence one connects ``stubs'' of various degrees completely at random, until each node has reached the number of links corresponding to its assigned degree. We have proposed in \cite{di2023generation} an object-based implementation of this procedure using Python-networkX. The degree sequence can be constructed from the theoretical degree distribution following different approaches; our preferred method is one, called ``random-hubs'' method, in which statistical fluctuations are minimized for all the degrees such that $P(k)N >1$ (i.e., those for which at least one node with that degree is present, on average).

We recall that the use of the configuration model networks \cite{newman2001random,newman2003structure} for the study of diffusion processes has been extensively investigated for the purpose of modeling SIR processes in \cite{serrano2007correlations,miller2011note}. The proposed graphs span a few tens of nodes, but a limit for large networks is also introduced, so that the procedure can be applied without creating multiple edges and self-loops. 

Condition (b) (assigned correlations) can be accomplished, with several limitations, via Newman rewiring \cite{newman2003mixing}. In the following  subsections we describe the Newman rewiring and the random-hubs method, with reference to our recent numerical implementations.

\subsection{Principle of the Newman rewiring}

The Newman rewiring requires a ``target'' correlation 
matrix as a function of the excess degrees of the nodes in the network. The links $(a,b)$ and $(c,d)$ chosen at random in the list of links,
are substituted with $(a,c)$ and $(b,d)$ with probability 1 if $E_2 > E_1$, where $E_1$ and $E_2$ depend on the elements of the target correlation matrix $e^0$ (see general definition of the matrix elements $e_{jk}$ in Sect.\ \ref{stat_descr}). More exactly, $E_1 = e^0_{AB} e^0_{CD}$ ($A$, $B$, $C$ and $D$ are the excess degrees of the nodes $a$, $b$, $c$ and $d$ involved in the rewiring). $E_2$ is the analogous quantity for the $(a,c)$ and $(b,d)$ links. If $E_2 \le E_1$, the rewiring is performed with a probability proportional to the ratio $E_1/E_2$. A further check on the presence of the edges is introduced in order to avoid repetitions.
In case of repetitions the rewiring must be skipped, or the exact degree sequence will not be preserved.

Section II of Ref.\ \cite{di2023generation}, describes a Newman rewiring procedure which takes advantage of the power of the Python library NetworkX and starts from a  random network
with given probability parameter. The \texttt{random\_reference} function actually constitutes the core of the procedure and makes use of the \texttt{degree\_dist} function, which returns the normalized frequency for the distribution of degree values, in order to obtain the probability sequence of nodes. This functions checks the usual conditions for random rewiring:
all vertices should be different and no parallel edges must be created in the network. 

Typically a few hundreds of rewiring cycles are needed for a better statistics, even if after a few cycles there is already convergence to the target. The $\bar{k}_{nn}(k)$ function for the target correlation matrix is also computed for comparison with the average $\langle \bar{k}_{nn}(k) \rangle$ of the rewired ensemble. The $\langle \bar{k}_{nn}(k) \rangle$ in \cite{di2023generation} shows a typical assortative behaviour, being represented by an increasing function for all values of the degree $k$ except for the largest ones, for which a structural disassortativity is observed, namely the hubs are hardly connected to other hubs. 

The generation of random scale-free networks and consequent manipulation has been important in the tradition of literature on social network analysis \cite{4658141} and on protein networks, which have been compared to networks following a power law \cite{fadhal2014protein}. The importance of building a scale-free network with an assortative matrix for the purpose of studying the Bass diffusion model was addressed in \cite{bertotti2019evaluation}. In that context, the exponent of the power law was varied between 2 and 3 and correspondingly the diffusion peak on the network was found.  

These results were reproduced and extended in \cite{di2023generation} introducing scale-free networks with power law exponent between 2 and 3 in the configuration model construction, namely inserting as input a network constructed with the random hubs method and a given exponent, as shown in Section IV of \cite{di2023generation}. 

\subsection{Random hubs and assortative matrices}
\label{random-hubs}

\begin{figure}[ht] 
\centering 
\includegraphics[width=0.8\columnwidth]{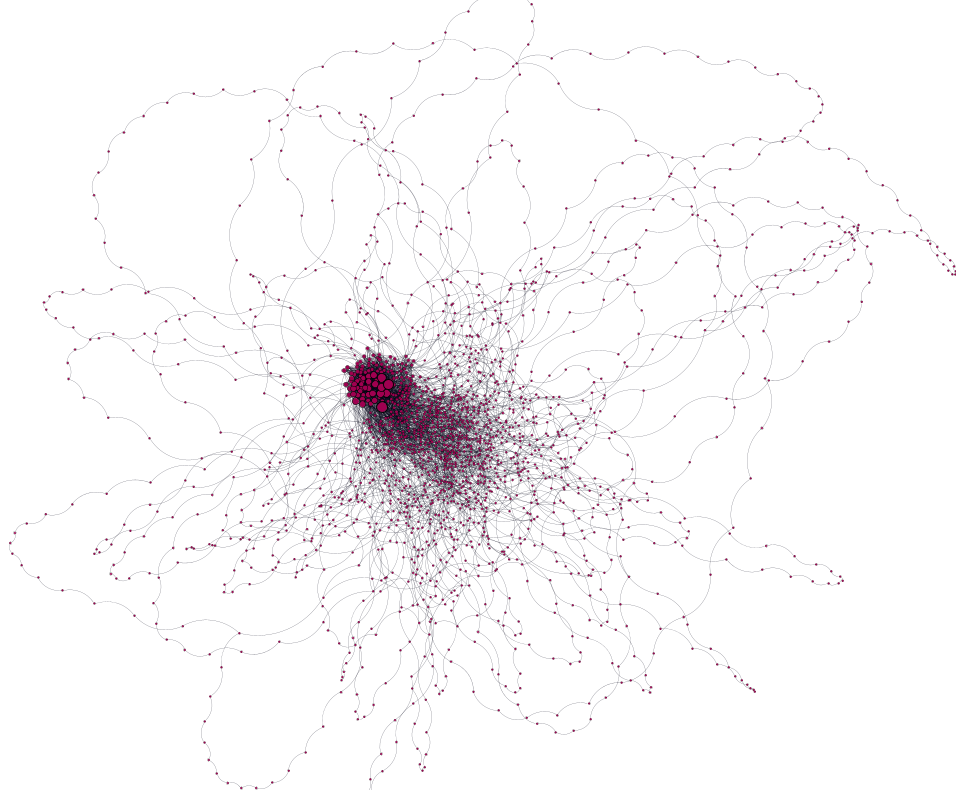}
\caption{Example of assortative scale-free network with $\gamma=2.5$, $k_{min}=2$, 2000 nodes, built via Newman rewiring with target correlation BM1 (Sect.\ \ref{random-hubs}).
Note the directly interconnected hubs and the long chains of nodes with degree 2.
}
\label{MAX--ass}
\end{figure}

The random hubs method allows to construct networks that are close to the ideal scale-free graph and works as follows. If the average number of nodes
with degree $k$ is smaller than 1, i.e., $NP(k) = X < 1$, then a node with this degree
will be created with probability $X$. In order to cover  all
the foreseen degrees, a random variable $\xi \in (0, 1)$ is generated for each value of $k$. Then denoting by $Int(NP(k))$ the integer part of $NP(k)$ and by $Dec(NP(k))$  its decimal part, one sets the number $N_k$ of nodes with degree $k$ to
$N_k = Int(NP(k))$ if $\xi > Dec(NP(k))$ and $N_k = Int(NP(k)) + 1$ if $\xi < Dec(NP(k))$. This technique can be combined with the assortative rewiring; however, one must decide which kind of target works better for generating an assortativity that is definitely greater than 0. In particular two classes of assortative matrices called BM1 and BM2 have proven to be effective in generating a coefficient $r \approx 0.2 \div 0.3$. The assortative matrix BM1 was applied in \cite{di2023generation} to an uncorrelated network of 10000 nodes with $\gamma =2.5$, obtaining an ensemble of 400 rewired networks with Newman coefficient $r=0.337 \pm 0.002$. In each of the rewiring subcycles, the number of accepted assortative rewiring steps was $1.45 \times 10^5$. (An example with 2000 nodes is shown in Fig.\ \ref{MAX--ass}.) The BM2 matrix, which is related to the excess degrees in the network, yielded a slightly less efficient process, guaranteeing a Newman coefficient of $r=0.176 \pm 0.005$ on an ensemble of 200 networks obtained with subcycles in which the number of accepted rewiring steps was $2.8 \times 10^5$. The methods for constructing the assortative matrices are explained in \cite{bertotti2019evaluation}. 

\section{Disassortative rewiring}
\label{dis_rew_analysis}

\subsection{Newman rewiring}
\label{newman-rew}

The importance of performing a Newman rewiring in order to build \emph{assortative} networks has already been stressed in \cite{bertotti2020network} and \cite{di2023generation}, where it was recognized that it allows recreating the dynamics for diffusion processes in networks.

\begin{figure}[ht] 
\centering 
\includegraphics[width=0.5\columnwidth]{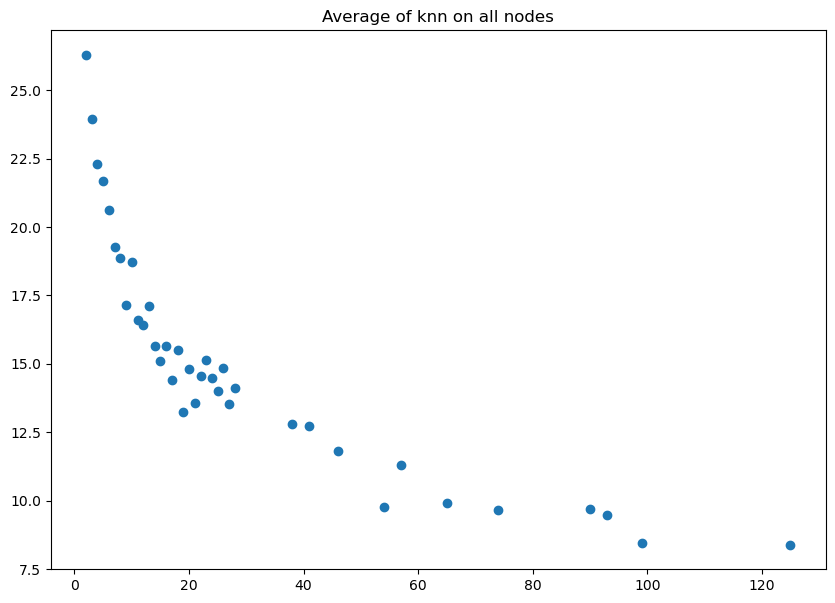}
\includegraphics[width=0.5\columnwidth]{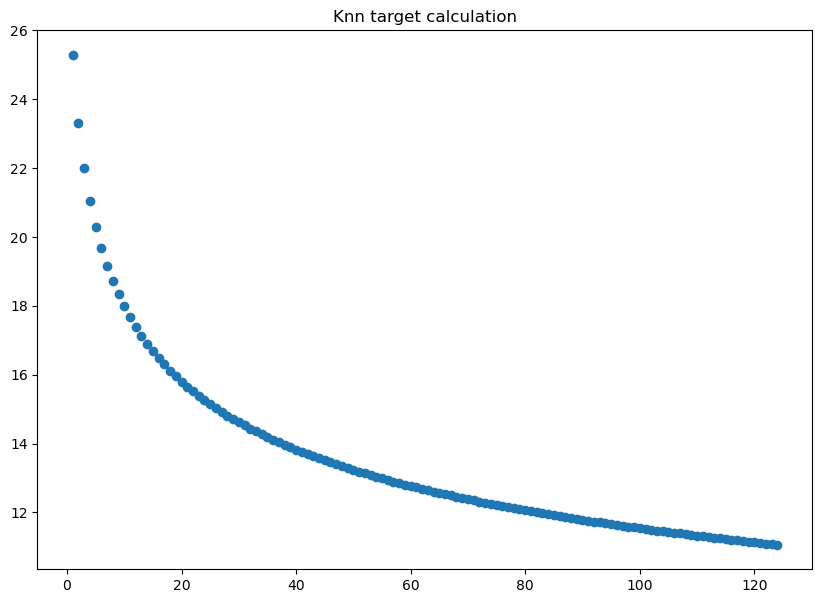}
\caption{
The graphics shows the scatter plot for the average $\bar{k}_{nn}$ and the target $\bar{k}_{nn}$ for a scale-free network after two cycles with 70000 disassortative rewirings by means of the Porto-Weber recipe (horizontal axis: node degree). The network is a scale-free with $\gamma = 2.5$ and  1000 nodes; it is disassortative with $r=-0.15$. 
}
\label{fig-dis-Porto}
\end{figure}

With the same code used in \cite{di2023generation} it is possible to perform a Newman rewiring with disassortative target correlations of the kind obtained with the Porto-Weber method (Sect.\ \ref{bass-model}).  A procedure for the construction of ensembles of scale-free networks which employs the Porto-Weber correlation matrices has been already presented in \cite{silva2019spectral}. This procedure uses a modified version of the configuration model, adapted to the Porto-Weber matrices.
Compared to that work, in our case the standard configuration model with the \texttt{Porto} function for reconstructing $P(h\lvert k)$ was used, which meant using a more concise code, very fast in the running phase -- a few seconds on an Apple M1 chip with 8 GB memory. The miminum degree was set to $k_{min}=2$, the network had 1000 nodes and  a scale-free structure with $\gamma=2.5$. The coefficient $\beta $ was chosen such that $\bar{k}_{nn}\propto k^{-0.2}$. After the rewirings with the Newman method including the Porto Weber recipe as an input, the target $\bar{k}_{nn}$ was computed for comparison starting from the $P(h\lvert k)$ with the sum $\bar{k}_{nn}=\sum_h hP(h\lvert k)$. The results are presented in Fig. \ref{fig-dis-Porto}. The $\bar{k}_{nn}$ shows the predicted decreasing behavior for the disassortative case.

\subsection{The formula for $\Delta r$ in a rewiring step, and relation to $\Delta K$}
\label{relation}


As a premise for the next sub-section, which is about maximally disassortative rewiring, it is necessary to recall the formula that gives the variation of the Newman assortativity coefficient $r$ in one degree-conserving rewiring step \cite{menche2010asymptotic,bertotti2019configuration}. The formula is
\begin{equation}
    \Delta r = \frac{2(-AB-CD+AC+BD)}{L\sigma_q^2}
\end{equation}
where $r$ is the Newman assortativity coefficient, $L$ is the number of links and $\sigma_q^2$ is the same quantity that appears in the definition of $r$ (Sect.\ \ref{stat_descr}). $A$, $B$, $C$ and $D$ are the excess degrees of the nodes $a$, $b$, $c$ and $d$ involved in the rewiring. In Ref.\ \cite{bertotti2020network} an interesting connection was proven between the variation of $r$ and the variation of a quantity which a priori does not appear to be directly related to $r$, namely $K$, the network average of the degrees of the first neighbors of each node. $K$ is a crucial quantity for predicting epidemic diffusion on scale-free networks. In \cite{boguna2003epidemic} it is proven that as a function of the network size $N$, $K$ diverges when $N\to \infty$, and this implies that phenomena of epidemic diffusion always propagate to the entire network, no matter how small the contagion probability. The said connection is
\begin{equation}
    \Delta K = -\frac{L\sigma_q^2}{2N} \, \frac{(AD+BC)}{ABCD} \Delta r
\end{equation}

\begin{figure}[ht] 
\centering \includegraphics[width=0.8\columnwidth]{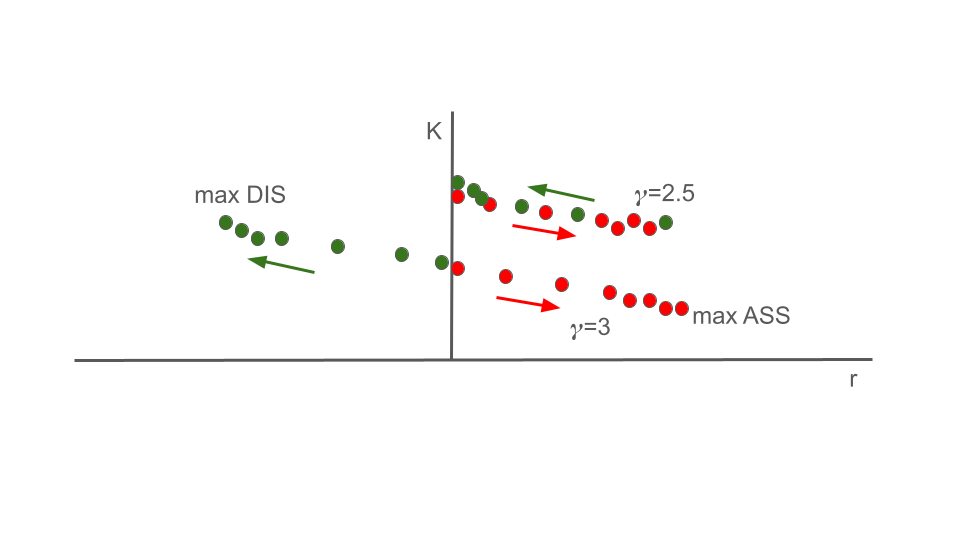}
\caption{
Typical situations in the maximally assortative and disassortative rewiring of scale-free networks, represented in the $r$-$K$ plane. In red, assortative trajectories, in green disassortative trajectories. At the end of the trajectories, the points representing the networks become more dense (see real data in \cite{bertotti2020network}). The $K$-intercept depends on $n$ and $\gamma$, being larger for smaller $\gamma$. The trajectories are reversible, as shown for the trajectory with $\gamma=2.5$.
}
\label{ass-dis-rew}
\end{figure}

Due to this relation, the variations of $r$ and $K$ are always opposite in sign and the rewiring trajectories in an $r$-$K$ plane  always have negative slope. In this work, we are focusing our attention on maximally disassortative rewiring cycles, therefore on trajectories which starting close to the vertical axis $r=0$ move to the left and increase from right to left (Fig.\ \ref{ass-dis-rew}). In \cite{bertotti2020network} we also have described the opposite kind of trajectories, leading to maximally assortative networks, whose features are obviously very different. The new code that we have implemented for this work using Python-NetworkX allows to check easily that the rewiring trajectories are reversible, i.e., upon changing in the algorithm the criterion $\Delta r>0$ into $\Delta r<0$ and vice-versa, their direction is inverted but without any ``hysteresis'' in the $r$-$K$ plane. This implies, at least at a numerical-heuristic level, that not only a differential relation between $r$ and $K$ exists, but this relation can be integrated in a path-independent way to obtain, at fixed $\gamma$ and $n$, a complete relation $r(K)$ or $K(r)$.

\begin{figure}[ht] 
\centering 
\includegraphics[width=0.7\columnwidth]{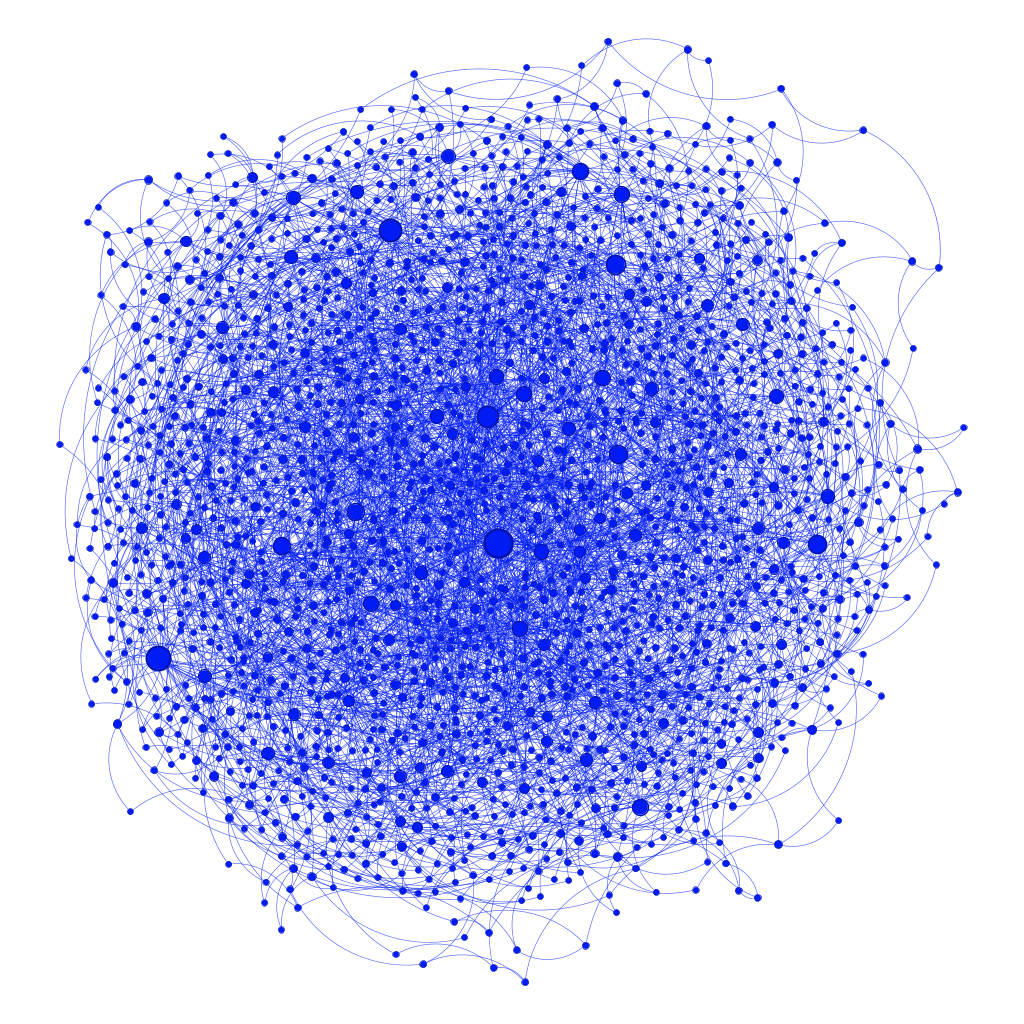}
\caption{Example of maximally disassortative scale-free network with $\gamma=2.5$, $k_{min}=2$, 2000 nodes, built via a rewiring based on the condition $\Delta r<0$ (Sect.\ \ref{maximally}). The hubs are connected through nodes of small degree, compare Figs.\ \ref{hub_dis}, \ref{hist_hub_dis}.
}
\label{MAX--dis}
\end{figure}

\begin{figure}[ht] 
\centering \includegraphics[width=0.6\columnwidth]{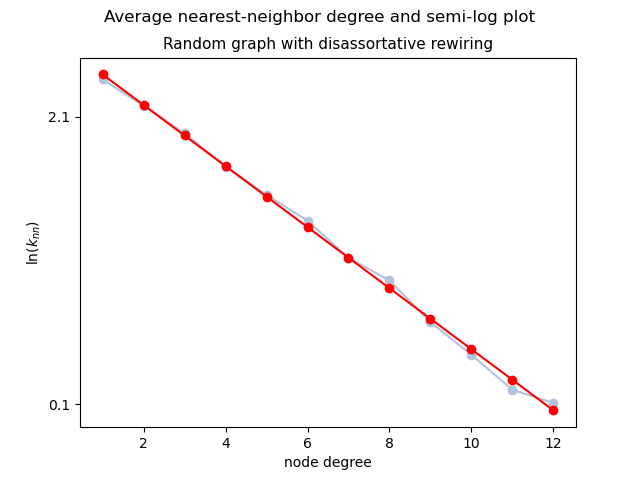}
\caption{
The $\ln(\bar{k}_{nn})$ function of a random network with 4000 nodes and probability $p=0.001$ after one cycle of maximally disassortative rewiring . The Newman coefficient of the ensemble is $r=-0.94 \pm 0.003$. The giant component after the last rewiring is $99.7 \%$. In the rewiring cycle the number of accepted rewiring steps was about 10'000. In red: linear fit.
}
\end{figure}

\subsection{Maximally disassortative rewiring based on the criterion $\Delta r<0$}
\label{maximally}

In this case, a rewiring step is always accepted if it causes a variation $\Delta r<0$, and accepted if $\Delta r>0$ with a small return probability depending on a tunable ``temperature'' $T$. Starting from an uncorrelated network with degree distribution of the random, Erd\"os-Renyi type, one generally obtains in an efficient way a maximally disassortative network with $r$ quite close to $-1$. For example, with a random network of 4000 nodes and a wiring probability equal to 0.001, which turns out in a certain realization to have $k_{max}=12$ and $\langle k \rangle=4.04$, after a single rewiring cycle with ca.\ $10^4$ accepted steps one obtains $r=-0.94$. One observes, quite unexpectedly based on the standard references (\cite{barabasi2016network} and refs.), that the resulting $\bar{k}_{nn}$ is well fitted by a linear plot in semi-log scale, i.e., it is of the form $\bar{k}_{nn}\simeq a \exp (bx)$, with $\ln a=2.62$, $b=-0.22$. In a log-log scale, supposing as usual a power law $\bar{k}_{nn}\simeq ck^\mu$, one obtains formally $\mu=-0.20$, even though the agreement is clearly worse. This implies that in a $r-\mu$ plot the network is placed well on the left of the bisector with slope 1. For the $r-\mu$ plots in general see the book of Barabasi \cite{barabasi2016network}, Ch.\ 7, also freely available online.

Starting from an uncorrelated scale-free network, a maximally disassortative rewiring gives networks with absolute values of $r$ definitely smaller than 1, as already reported in \cite{bertotti2020network}. For example, with $\gamma=2.5$, $k_{min}=2$, $N=2000$ one obtains in a certain realization $k_{max}=107$, $\langle k \rangle=4.29$ and $r=-0.34$. Both the semi-log and log-log fits show a poor agreement. The slope of the log-log fit is $\mu \simeq -0.35$, therefore in a $r-\mu$ plot the network is placed close to the bisector with slope 1, like most of the real disassortative scale-free networks \cite{barabasi2016network}. Note that also the network of Fig.\ \ref{fig-dis-Porto} is close to the bisector in an $r-\mu$ plot, because its $\bar{k}_{nn}(k)$ is well approximated by $\sim k^{-0.2}$, and $r=-0.15$.

\begin{figure}[ht] 
\centering \includegraphics[width=0.6\columnwidth]{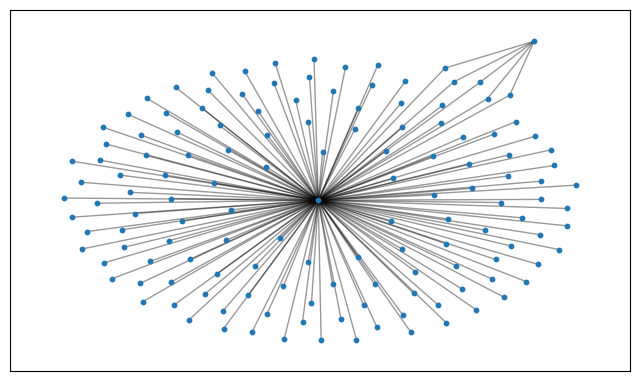}
\caption{
The largest hub of a typical maximally disassortative scale-free network with $k_{min}=2$, plus all its neighbors and one of the second neighbors. The strong disassortativity implies that all first neighbors have degree 2 and the second neighbors have degree definitely greater than 2 (compare histogram in Fig.\ \ref{hist_hub_dis}).
}
\label{hub_dis}
\end{figure}

\begin{figure}[ht] 
\centering \includegraphics[width=0.6\columnwidth]{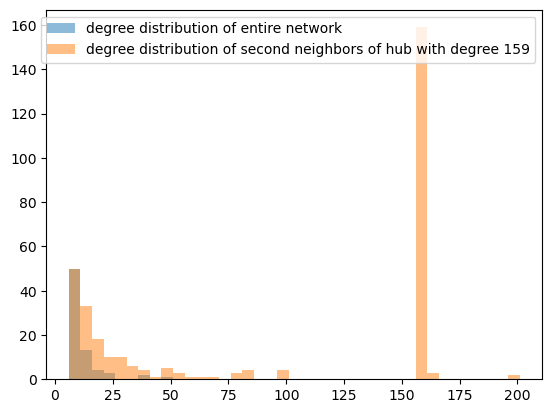}
\caption{
Histogram of the degree distribution of a maximally disassortative scale-free network (dark) superposed to the histogram of the degree distribution of the second neighbors of the largest hub. While the degree distribution of the full network goes quickly to zero, that of second neighbors reaches a minimum and then increases, due to the strong disassortativity. The spike at degree 159 occurs because the hub itself has been counted among the second neighbors.
}
\label{hist_hub_dis}
\end{figure}

The topological structure of maximally disassortative scale-free networks is quite interesting. In the previously studied case of $k_{min}=1$ \cite{bertotti2020network,moreno2003disease} the networks turned out to be completely fragmented, with the largest connected components all having the form of isolated ``stars'', in which a hub is connected only to nodes of degree 1, as enforced by the disassortativity condition. In the case with $k_{min}=2$, almost all hubs are exclusively surrounded by first neighbors of degree 2. (It is easy to verify this just checking when the values of $\bar{k}_{nn}$ are exactly equal to 2.) Large groups of these first neighbors tend to connect with other hubs or to nodes of intermediate degree (see Fig.\ \ref{hub_dis}), thus forming a strongly connected structure which does not resemble either a tree or a maximally assortative network (Fig.\ \ref{MAX--dis}). In order to better analyze this property one can, chosen a large hub having only first neighbors of degree 2, create an histogram of the degrees of all its second neighbors; it is easy to verify that all first neighbors have degree 2 by checking if $\bar{k}_{nn}(k)=2$ exactly when $k$ is the degree of the hub. The histogram shows (Fig.\ \ref{hist_hub_dis}) that no neighbors of degree 2 or 3 are present, then there are some of degree 4, 5 etc., but while the degree distribution of the whole network goes quickly to zero after these values, the degree distribution of the second neighbors reaches a minimum and then increases, clearly because of the strong disassortativity.

\

\section{Agent-based models}
\label{abm}

In \cite{di2024bass} the assortative scale-free networks generated with the codes of \cite{di2023generation} were exported in \texttt{graphml} format into the free simulation software NetLogo, for running agent-based simulations of the Bass diffusion model. The Python connector PyNetLogo was used to perform numerical and statistical analyses of the simulation results. Information obtained from the agent-based simulations about peak diffusion times can be compared with the results of models based upon differential equations. The results are summarized in the Conclusions of this work.

A strong correlation was also found to occur, averaging over ensembles of assortative scale-free networks, between the function $\langle \bar{k}_{nn}(k) \rangle$ and the function $\langle \bar{C}(k) \rangle$, which represents the mean clustering coefficient of nodes of degree $k$ .

\subsection{Simulations of signed networks with NetLogo}
\label{signed}

There exists a significant series of studies on signed networks after the classic work by \cite{harary1953notion} about the socio-psycological interpretation of positive and negative relationships between groups of individuals; many of the addressed subjects, such as the proportion and the weight of the negative links, are taken into account e.g. in the recent paper by Kirkley et al. \cite{kirkley2019balance}. This series of analyses is based on the concept of balanced networks, in which the signs of the links are not random but can be determined starting from a study of the adjacency matrix of the graph. It was stressed in \cite{estrada2014walk} that, although balanced structures can be constructed for theoretical social networks, they do not reflect the situation of the real online social networks such as internet databases for voting and evaluating items' quality. 

Based on this assumption, Mueller and Ramkumar \cite{mueller2023signed} have decided to model the networks describing the diffusion of innovations by assigning positive and negative signs that are randomly distributed with probability $p_n$. They introduce a threshold condition for the adoption of the innovation: the sum $\theta_{-}*f_{FOE} + \theta_{+} $ must be strictly positive. Here $\theta_{-}$ represents the number of first neighbours who have already adopted and are connected to the agent through a negative link, and similarly for $\theta_{+}$. The $f_{FOE}$ parameter allows to define the strength of the influence of each negative link and consequently to combine the concept of threshold for adoption with the picture of a signed network. 

In recent times NetLogo has been quite efficiently used for studies in sociological behaviour in large groups, see e.g. \cite{sznajd2024toward}. In this sense, it is not surprising that we chose to use it among other platforms  \cite{railsback2006agent} to simulate the dynamics of the adoption in signed networks as the ones planned by Mueller and Ramkumar. The parameters $p_n$ and $f_{FOE}$ and the kind of network were varied. The starting scenario is a random network with 300 nodes, 900 links and a population of 3\% initial adopters randomly chosen. This scenario was then further extended to networks generated with networkX using the configuration model and Newman rewiring. These networks are imported into NetLogo, and then NetLogo performs the agent-based simulations using the threshold transition rule of \cite{mueller2023signed}. The principal output of the simulations is the final percentage of adopters. While in the traditional Bass model the entire population eventually adopts the innovation, here the presence of a threshold has the consequence that a certain number of individuals never adopt, especially if their neighbours are connected with negative links.

\section{Conclusions}
\label{conc}

The analysis of diffusion times for the Bass model on scale-free networks with the standard initial conditions (no adopters at time $t=0$) in dependence on degree correlations has shown that the peak time is definitely smaller for assortative networks, compared to uncorrelated networks. This holds for real networks and has been verified both through the solution of the differential equations for the single nodes and through agent-based simulations. In the mean-field approximation (HMF) the opposite is found, and this gives a clear example of the limits of that approximation. 

Both in the solutions of differential equations and in the agent-based simulations one observes a strong dependence of the peak times on the degree of the largest node effectively present in the network. When this degree is very large (i.e., in the presence of a giant hub), $t_{max}$ is markedly smaller. A dependence of this kind cannot clearly be captured by the HMF approximation, in which all hubs are virtually present, albeit with a very small probability. It is also clear that in mean-field it is impossible to distinguish, e.g., between the adoption time of small nodes in a long assortative chain (and there are many such chains in a real assortative network) and the adoption time of nodes of the same degree placed close to the hubs, which tend to adopt much earlier.

For the construction of ensembles of real \emph{disassortative} networks we have given in this work some new preliminary results. Agent-based simulations have not been run yet, but the solutions of differential equations with single nodes show that diffusion is in that case slightly slower than with uncorrelated networks. This is in agreement with the HMF approximation.

In general, the Newman rewiring is confirmed to be a powerful method for the construction of statistical ensembles of networks having pre-assigned degree correlations, both in the case of random networks and scale-free networks. Further variations, like e.g.\ small-world networks or scale-free networks with degree distribution modified at large or small degrees can be quickly tested using our codes which include both the Configuration Model and Newman rewiring.

The maximally assortative/disassortative rewiring method is useful for exploring parameter spaces of networks and seeing what regions can actually be reached. In particular, we chose to do this for the $r$-$K$ plane, because there turns out to be a general differential relation between the variations $\Delta r$ and $\Delta K$ in any degree-conserving rewiring. From the numerical trials, the differential relation appears to be reversible, i.e.\ integrable. We recall that $K$ (network average of the degrees of the first neighbors of each node) is closely related to the epidemic threshold, because it can be shown that when $K$ grows with the networks size $N$, like in scale-free networks, the threshold value of the contagion probability for epidemic diffusion to the entire network tends to zero.

Finally, for signed networks our agent-based simulations confirm and extend the results by Mueller-Ramkumar \cite{ramkumar2022diffusion}, in dependence on the parameters $f_{foe}$ (strength of influence of negative links) and $p_n$ (probability of negative links). In several cases, diffusion to the entire network is blocked by negative links, especially in the presence of high clustering (which for the hubs, as we have shown, is strongly correlated with assortativity).

\bigskip

\bigskip
\noindent \textbf{Acknowledgments -} This work was financially supported by the Free University of Bolzano-Bozen with the research project NMCSYS-TN2815. G.M.\ is a member of INdAM (Istituto Nazionale di Alta Matematica). 

\bigskip
\noindent \textbf{Competing interests statement -} The authors declare that there are no relevant financial or non-financial competing interests to report.

\bigskip
\noindent \textbf{Data availability statement -} The authors declare that the manuscript has no associated data. 

\bibliographystyle{ieeetr}
\bibliography{nets}

\end{document}